\documentclass[twocolumn,showpacs]{revtex4}

\usepackage{amsmath}
\usepackage{amssymb}
\usepackage{epsfig}

\newcommand{\beq}{\begin{equation}}
\newcommand{\eeq}{\end{equation}\noindent}
\newcommand{\bean}{\begin{eqnarray*}}
\newcommand{\eean}{\end{eqnarray*}\noindent}
\newcommand{\bea}{\begin{eqnarray}}
\newcommand{\eea}{\end{eqnarray}\noindent}

\begin{document}

\title{Description
of Heavy Quark Systems by means of Energy Dependent Potentials}

\author{R.J. Lombard$^1$, J. Mare\v{s}$^2$ and C. Volpe$^1$}
\email{lombard@ipno.in2p3.fr,mares@ujf.cas.cz,volpe@ipno.in2p3.fr} 
\affiliation{$^1$ Institut de Physique Nucl\'eaire, F-91406 Orsay cedex, 
France \\
 $^2$ Nuclear Physics Institute, 25068 \v{R}e\v{z}, Czech Republic}

\date{\today}

\begin{abstract}
We apply, for the first time,
an energy dependent Schr\"odinger equation to describe
static properties of heavy quark systems, i.e. charmonium and bottonium.
We show that a good description of the eigenstates and reasonable
values for the widths can be obtained. Values of the radii and of 
the density at the origin are
also given. We compare the results to those
deduced with a Schr\"odinger equation implemented with potentials used 
so far.  We note that the energy dependence of
the confining potential provides a natural
mechanism for the saturation of the spectra. Our results introduce a new 
class of potentials for the
description of heavy quark systems.

\end{abstract}

\maketitle
\noindent

Wave equations with energy dependent potentials are already familiar in
physics.
The Pauli-Schr\"odinger equation  describing a particle in an external
electromagnetic field \cite{pauli} or the Hamiltonian formulation of
relativistic quantum mechanics, in connection with manifestly covariant
formalism with constraints \cite{hagop,crater,mourad}, offer two examples of energy dependent
potentials.
Such dependence arises from the relativistic nature of the problem. 
Moreover, it has been shown that in the context of relativistic quantum
mechanics, the potentials can be derived from fundamental theories, i.e 
a link exists with field theory \cite{hagop3}.
The present work has a different aim, namely to show the
characteristic features which can be expected from energy dependent
potentials, when applied to a specific physical system. 
For this purpose, simple
potentials admitting analytic or semi-analytic solutions are of great help
since they facilitate the comparison with  usual potentials.  

The introduction of an energy dependence has several 
implications with respect
to usual quantum mechanics. For example, the conservation of the norm
asks for the modification of the scalar product \cite{hagop2}.
The energy dependent potentials must fulfill certain conditions 
in order to result in a meaningful quantum theory \cite{flm}. 
For those potentials which are acceptable, the corresponding 
Schr\"odinger equation is equivalent to a Schr\"odinger equation 
including a non-local potential. The non-locality, however, is 
treated more efficiently under the form of an energy dependent potential.
A detailed analysis of these formal aspects is performed in \cite{flm}
showing that the properties of a good quantum theory are well preserved.

The scope of this letter is to apply, for the first time, such an
approach to cases of physical interests. 
A natural application is offered by heavy quark systems.
Their description has been a quite
successful playground for potential models 
in the past \cite{BT,cornell,log,powerlaw,EQ,brambilla}.
Here we show that 
energy dependent potentials represent a new class of potentials
that can account for the properties of
charmonium and bottonium. 
In particular, we present the eigenstates of these systems, the
root mean square radii, the leptonic widths and the density of the S-wave functions 
at the origin, the latter being a useful input for quarkonium production in
high-energy collisions.
We compare our results to those obtained
with the potentials known so far
to give reasonable description of charmonium and bottonium properties. 
We take the Buchm\"uller and Tye  ({\sc BT}) \cite{BT} and the Cornell \cite{cornell}
potentials as examples. 

For simplicity we consider spherical symmetric potentials, assuming a linear
energy dependence.
The corresponding Schr\"odinger equation
reads (ignoring spin degrees of freedom):

\begin{equation}\label{1}
[ - \frac{\hbar^2}{2m}   \vec{\nabla}^2 + V(r,E) ]
\Psi_{n,\ell}(\vec{r})= E_{n,\ell}\Psi_{n,\ell}(\vec{r}) ,
\end{equation}
\noindent
where the potential used is:
\begin{equation}\label{2}
V(r,E) = (1 + \gamma E)f(r) + D_0 \vec{\ell}^2 + V_0.
\end{equation}
\noindent
with $n,\ell$ denoting the principal and the
orbital quantum numbers respectively and $m/2$ the reduced mass
(we work in units $\hbar$=197.3~MeV fm$^{-1}$).
To illustrate our aim we choose two
radial dependences for which analytical solutions exist:
{\sc i)} the linear one, i.e.
$f(r)= \lambda m r$ and {\sc ii)} the harmonic one, i.e.
$f(r)= m\omega^2 r^2/2$. In each case the total number of parameters is 5.
Note that only four of the parameters are
free since $M(q\bar{q})= 2m+V_0$.
They are determined by constraining
the energy differences between the ground state and the 1P, 2S, 2P and 3S
states to the experimental energies for both systems. The 1D state is also
considered for the $c\bar{c}$ system.
Note that for the two chosen radial dependence, only negative values of
$\gamma$ are acceptable \cite{flm}.

The term $D_0 \vec{\ell}^2$ is added to break the degeneracy between
the states of different angular momentum. Its contribution is  
necessary to achieve accurate fits, especially in the case of 
the harmonic oscillator.   

In the harmonic dependence case, the energies are obtained by solving 
the following quadratic
equation:
\begin{eqnarray}
E_{n,\ell}^2 & -&  \Big[ 2D\ell(\ell+1) + a \hbar^2
\omega^2 \gamma \Big] E_{n,\ell} \\ \nonumber
& + &  D^2\ell^2(\ell +1)^2 - a
\hbar^2\omega^2  = 0 \,
\end{eqnarray}
where $D = \frac{\hbar^2}{2m} D_0$ and
$a=(4n + 2\ell +3)^2/4 $; while for the linear dependence
the S-state eigenvalues are obtained by solving
\begin{equation}
(E_{n,0} + V_0)^{3/2} - A_n \sqrt{m} \lambda (1 + \gamma E_{n,0}) = 0 \ , 
\end{equation}
with $A_n = (-a_n)^{3/2} \frac{\hbar}{\sqrt{2}}$, $a_n$ being the n-th zero
of the Airy function.

Eqs. (3) and (4) are non-linear. Therefore one has to specify the choice of
the eigenvalue in the case when several solutions exist. For the harmonic 
oscillator,
only positive energies lead to square integrable wave functions (see
\cite{flm}). Consequently negative solutions are rejected. For the linear
potential, Eq. (4) can be transformed into a cubic equation, which has a
single real root in the range of parameters used in the present work.

In the framework of potential models, the leptonic widths
of the S-states, without radiative and relativistic corrections
\cite{rad1,rad2,rad3}, 
are given by the van Royen-Weisskopf formula \cite{weiss}: 
\beq\label{weiss}
\Gamma_{ee}(nS) = {\frac{16 \pi e_{q}^2 \alpha^2} {M^2(q \bar{q})}}
|R_{n,0}(0)|^2
\eeq
where $R_{n,0}$ indicates the radial component of the wave function
evaluated at the origin, $e_q$ the quark charge and $M(q\bar{q})$
is the physical mass.
We have checked that the same formula holds when using energy dependent 
potentials.

Table I shows the sensitivity of the results to the variation of the
parameters, in the case of the harmonic oscillator radial dependence
as an example. The remarkable fact is the change in the spectrum pattern
brought by the energy dependence, which deepens the ground state
with respect to the other states.
The usual harmonic oscillator would be
unable to produce the relative positions of the 1P, 2S, 1D and 3S levels, 
even in the presence of the $\vec{\ell}^2$ term in the Hamiltonian.
On the other hand, the energy dependent potential with 
the harmonic oscillator radial shape is not able 
to account for both the 1D and  3S states simultaneously. 
In this respect, 
the linear potential comes out
to give much better description of the $c\bar{c}$
and $b\bar{b}$ properties.

\begin{table}[ht]
\begin{tabular}{|c|cccc|c| }
\hline \hline
{\sc spectrum} & & & & & {\sc exp.} \\
 1P &  397 & 397 & 397 & 397 & 397 \\
  2S & 589 & 589 & 589 & 589 & 589 \\
  1D & 758.5 &  759.2 & 706.2 & 706.4 & 710     \\
  2P & 839.0 & 839.6 & 789.5 &  789.6 & 840  \\
  3S & 946.2 & 947.1 & 876.7 & 876.8  & 940   \\
  & & & & & \\
  {\sc w.f. properties}  & & & & & \\
  $\Gamma_{1 {\rm S}}$ & 2.6 & 6.2 & 2.9 & 7.7  & 5.26(.37)  \\
  $\Gamma_{1 {\rm S}}/\Gamma_{2 {\rm S}}$ & 1.7 & 1.5 & 2.3 & 1.7 & 2.4 \\
  $ \langle r^2 \rangle_{1 {\rm S}}^{1/2}$ & 0.490 & 0.374 &  0.469  
  & 0.346 &  \\
  & & & & &  \\
  \hline
  {\sc parameters} & & & &  &  \\
  $\omega~$(fm$^{-1}$)~~ & 2.7 & 2.3 & 3.5  & 3.0 &   \\
  $V_0~$(MeV)~~ & 0.0 & 600. & 0.0 & 600. &   \\
  $D_0~$(MeV)~~ &  2.8 &  4.3 &  2.4 & 3.6 &  \\
  $ \gamma \cdot 10^{-4}~$(MeV$^{-1}$) & -4.29 & -5.72 &  -4.75  &  -6.64 &   \\
  $m $(GeV) & ~1.212 & 1.812 & 1.138 &   1.738  &  \\
  & & & &  &  \\
  \hline \hline
  \end{tabular}
  \caption{{\sc Sensitivity of the $c\bar{c}$ properties
  to variations of the parameters:}
  The results display  the energy differences (MeV) with respect to the
  ground state, the leptonic widths
  (keV), the root mean square radius (fm) in the case of
  the harmonic oscillator radial dependence.
  Experimental data are given for comparison, when available.}
  \end{table}
\begin{table}[!ht]
\begin{tabular}{|c|ccc|c|}
\hline \hline
 & This work & {\sc BT} & {\sc Cornell} & {\sc exp.}   \\
 \hline
 & & & &  \\
 1P & 397 & 424 &  428   & 397  \\
 2S & 589 & 601 & 590 & 589 \\
  1D & 710 & 716 & 713   & 710~    \\
   2P & 840 & 872 & 871  &  840  \\
    3S & 943 & 1023 & 1015  & 940     \\
    $\Gamma_{1 {\rm S}}$ & 7.51 & 8.0 & 14.33  & 5.26(.37)   \\
    $~\Gamma_{1 {\rm S}}/\Gamma_{2 {\rm S}}$ & 2.18  & 2.17 & 2.22 & 2.4  \\
    $ \langle r^2 \rangle_{1 {\rm S}}^{1/2}$ & 0.33  &  0.42  & 0.47 &  \\
    $|R_{1 {\rm S}}(0)|^2$ & 0.761   &  0.810   & 1.454   &  \\
    & & & &  \\
    \hline \hline
    \end{tabular}
    \caption{{\sc $c\bar{c}$ properties}: Results obtained with an energy
    dependent
    potential and the linear
    radial dependence,
      in comparison with those of Buchm\"uller and Tye ({\sc BT}) \cite{BT} and of
    the Cornell \cite{cornell} potentials.
    The values listed correspond to the energies (MeV) with respect to the
    ground state, to the
    leptonic widths (keV), the root mean square radius (fm),
    the density at the origin (GeV$^3$).
    The parameters used are $\lambda$=2~fm$^{-1}$,$V_0$=1940~MeV, $D_0$=2~MeV,
              $\gamma$=-5.007~$10^{-4}$ MeV$^{-1}$ and $m$=~1913 MeV.
              Experimental data are also given.}
    \end{table}  
  
In Tables II and III we give the results, obtained with the linear dependence,
for the energies, the leptonic widths, the root mean square
radius, and the density of the 1S states at the origin.
We compare our results
with those obtained with previous potentials, in particular the {\sc BT}
 and the Cornell ones. Their values of the 1S energy density at the 
origin are taken from \cite{EQ}.
We see that a very good agreement on
the spectrum is obtained. On the other hand the width of the 1S state as
well as the ratio of the 2S to the 1S widths present the same quantitative
agreement with experiment as previous potentials.

\begin{table}
\begin{tabular}{|c|ccc|c|}
\hline \hline
 & This work & {\sc BT} & {\sc Cornell} & {\sc exp.}   \\
 \hline
 & & & &  \\
 1P & 426 & 430 & 498   & 428  \\
 2S & 563 & 560 & 591  & 563 \\
 1D & 798 & 680 & 747   &     \\
 2P & 818 & 790 & 852  &  792  \\
 3S & 895 & 890 & 936  & 895     \\
 4S & 1116  & 1160 & 1213  & 1120     \\
 $\Gamma_{1 {\rm S}}$ & 0.93   & 1.68 & 3.66 & 1.32(.05)   \\
 $~\Gamma_{1 {\rm S}}/\Gamma_{2 {\rm S}}$ & 2.73 & 2.2 & 2.17 & 2.54  \\
 $ \langle r^2 \rangle_{1 {\rm S}}^{1/2}$ &  0.23 &0.23 &  0.20  &  \\
 $|R_{1S}(0)|^2$ & 3.65  & 6.477  & 14.08  &  \\
 & & & &  \\
 \hline \hline
 \end{tabular}
 \caption{{\sc $b\bar{b}$ properties}:  Same as Table II, but here the results
 correspond to $\lambda$=0.6~fm$^{-1}$,$V_0$=2725~MeV, $D_0$=15~MeV,
 $\gamma$=-1.06~$10^{-3}$ MeV$^{-1}$ and $m$=~5475 MeV.}
 \end{table}
 
In this respect,
one should remember that
the van Royen-Weisskopf formula should be multiplied by a factor
(1-16$\alpha_s$/3$\pi$) corresponding to the (first-order) radiative
corrections \cite{BT,EQ}.
In \cite{mathiot1,mathiot2} it is shown that after the inclusion of
the relativistic corrections, potential models having 
$|R_{1S}(0)|^2$=0.8-1 (6-8)
GeV$^{3}$  for the J/$\psi$ ($\Upsilon$)
are able to reproduce the leptonic widths.
This point should be
reconsidered in the framework of energy dependent potentials. 
Here we shall merely compare the zeroth order values with the ones obtained
from other potentials. In the case of the $c\bar{c}$ system, the 1S
energy density at the origin is close to the value given by 
the {\sc BT} potential.
On the other hand, for the $b\bar{b}$ system, it is
sensibly lower than the BT value. This situation is quite strange, since the
value of $|R_{1{\rm S}}(0)|^2$ is closely related to the positions of the
S-states,
which are very well reproduced by the present model
(and is not due to a peculiar
set of parameters).
Indeed the density at the origin of the 1S-state
can hardly be increased by 10-15$~\%$
by choosing a lower coupling constant $\lambda$. Consequently, such a
behaviour may
point to the necessity of finding a more universal radial shape and/or 
of a more pertinent energy dependence. 

In conclusion, we show that a new class of potentials, namely
energy dependent potentials, give a good description of
charmonium and bottonium properties. We emphasize that
the energy dependence is rather naturally introduced, since it arises from
the relativistic aspect of the problem. As a starting point, we have chosen
potentials admitting analytical solutions, at least for the S states. This
is a great advantage in view of the high non-linearity of the differential
equation to be solved. However, 
it could be rewarding to study other radial shapes, as well as 
combinations of usual and energy dependent potentials. 
As far as the energy dependence is concerned, the linear one has the
advantage of simplicity in many respects as already emphasized in \cite{flm}. 

We would like to stress a very important feature of the energy dependence
for confining potentials, namely the saturation of the spectra. For usual
potential, $E_{n,\ell}$ increases regularly toward $\infty$ with $n$ or $\ell$.
In contrast for energy dependent confining potentials, $E_{n,\ell}$ tends
to a limit as $n$ and $\ell$ increase. Thus, the density of states becomes
rapidly very large, the splittings getting infinitesimally small, which
forbids the observation of individual states. We believe that this property  
could have relevant implications in many aspects of the 
quark-model physics. 

Finally, understanding the connection
between 
the phenomenological potentials used here  and more fundamental
theories of elementary particles deserves further studies. 
The application to other systems is foreseen.

\bigskip
{\small We thank Yuri Ivanov for providing us with useful information.}

\end{document}